\documentclass[aps,rpl,preprint,groupedaddress]{revtex4}
\begin{document}

\title{Entanglement teleportation via thermally entangled states of two-qubit Heisenberg XX chain}
\author{Ye Yeo}

\affiliation{Centre for Mathematical Sciences, Wilberforce Road, Cambridge CB3 0WB, United Kingdom}

\begin{abstract}
Recently, entanglement teleportation has been investigated in Phys. Rev. Lett. 84, 4236 (2000).  In this paper we study entanglement teleportation via two separate thermally entangled states of two-qubit Heisenberg XX chain.  We established the condition under which the parameters of the model have to satisfy in order to teleport entanglement.  The necessary minimum amount of thermal entanglement for some fixed strength of exchange coupling is a function of the magnetic field and temperature.
\end{abstract}

\maketitle

The linearity of quantum mechanics allows building of superposition states of a composite system
$S_{AB}$
that cannot be written as products of states of each subsystem
$(S_A$ and $S_B)$.
Such states are called entangled.  States which are not entangled are referred to as separable states.  An entangled composite system gives rise to nonlocal correlation between its subsystems that does not exist classically.  This nonlocal property enables the uses of local quantum operations and classical communication to teleport an unknown quantum state via a shared pair of entangled particles \cite{Bennett}.  In the standard teleportation protocol these local quantum operations consist of Bell measurements and Pauli rotations.  It is shown in \cite{Bose} that standard teleportation with an arbitrary entangled mixed state resource
$\chi_{AB}$
is equivalent to a generalized depolarizng channel
$\Lambda(\chi_{AB})$
with probabilites given by the maximally entangled components of the resource.  Quantum teleportation of single-body quantum state via single quantum channel shared between two parties has been studied by a number of authors (see references in \cite{Lee}).  In a recent paper \cite{Lee}, Lee and Kim considered teleportation of an entangled two-body pure spin-$\frac{1}{2}$ state via two independent, equally entangled, noisy quantum channels represented by Werner states \cite{Werner}.  In their two-qubit teleportation protocol, the joint measurement is decomposable into two independent Bell measurements and the unitary operation into two local one-qubit Pauli rotations.  They found that quantum entanglement of the spin-$\frac{1}{2}$ state is lost during the teleportation even when the channel has nonzero quantum entanglement, and in order to teleport quantum entanglement the quantum channel should possess a critical value of minimum entanglement.\\

Recently, the presence of entanglement in condensed-matter systems at finite temperatures has been investigated by a number of authors (see, e.g., \cite{Nielsen} and references therein).  The state of a typical condensed-matter system at thermal equilibrium (temperature $T$) is
$\chi = e^{-\beta H}/Z$
where $H$ is the Hamiltonian,
$Z = tr e^{-\beta H}$
is the partition function, and
$\beta = 1/kT$
where $k$ is Boltzmann's constant.  The entanglement associated with the thermal state $\chi$ is referred to as the thermal entanglement \cite{Arnesen}.  In \cite{Yeo}, quantum teleportation using the thermally entangled state of two-qubit Heisenberg XX chain as a quantum channel is considered.  In this paper, we investigate Lee and Kim's two-qubit teleportation protocol using two independent thermally entangled states of two-qubit Heisenberg XX chain.  In contrast to \cite{Lee}, we consider as input a two-qubit Werner state \cite{Werner}.  We find that quantum entanglement of the Werner state is lost during the teleportation even when the channel has nonzero thermal entanglement, in accordance with \cite{Lee}.  In order to teleport quantum entanglement, the parameters of the model have to satisfy (8).\\

The Hamiltonian $H$ for a two-qubit Heisenberg XX chain in an external magnetic field $B_m$ along the $z$ axis is
\begin{equation}
H = \frac{1}{2} J   \left(\sigma^1_A \otimes \sigma^1_B +
                          \sigma^2_A \otimes \sigma^2_B \right)+
    \frac{1}{2} B_m \left(\sigma^3_A \otimes \sigma^0_B +
                          \sigma^0_A \otimes \sigma^3_B \right)
\end{equation}
where
$\sigma^0_{\alpha}$
is the identity matrix and
$\sigma^i_{\alpha} (i=1,2,3)$
are the Pauli matrices at site
$\alpha = A, B$.
$J$ is real coupling constant for the spin interaction.  The chain is said to be antiferromagnetic for $J>0$ and ferromagnetic for $J<0$.  The eigenvalues and eigenvectors of $H$ are given by
$H |00\rangle = B_m |00\rangle$,
$H |\Psi^{\pm}\rangle = \pm J |\Psi^{\pm}\rangle$ and
$H |11\rangle = -B_m |11\rangle$, where
$|\Psi^{\pm}\rangle = \frac{1}{\sqrt{2}} (|01\rangle \pm |10\rangle)$.
For the system in equilibrium at temperature $T$, the density operator is
\begin{equation}
\chi_{AB} = \frac{1}{Z}
\left[
e^{-\beta B_m} |00\rangle \langle 00| +
e^{-\beta J} |\Psi^+\rangle \langle\Psi^+| +
e^{\beta J} |\Psi^-\rangle \langle\Psi^-| +
e^{\beta B_m} |11\rangle \langle 11|
\right]
\end{equation}
where the partition function
$Z = 2\cosh\beta B_m + 2 \cosh\beta J$,
the Boltzmann's constant $k \equiv 1$ from hereon and
$\beta = 1/T$.
To quantify the amount of entanglement associated with $\chi_{AB}$, we consider the measure of entanglement \cite{Lee},
$E(\chi_{AB}) = \max\{ -2 \sum_m \lambda_m^-, 0\}$ where
$\lambda_m^-$
is a negative eigenvalue of
$\chi_{AB}^{T_B}$,
the partial transposition of $\chi_{AB}$.
The density operator $\chi_{AB}$ is entangled if and only if $\chi_{AB}^{T_B}$ has any negative eigenvalues \cite{Peres, Horodecki}.
After some straightforward algebra, the amount of thermal entanglement is
\begin{equation}
E(\chi_{AB}) = \max\left\{
\frac{\sqrt{\cosh^2\beta B_m + \cosh^2\beta J - 2} - \cosh\beta B_m}{\cosh\beta B_m + \cosh\beta J}, 0
\right\}
\end{equation}
The amount of thermal entanglement is invariant under the substitutions
$B_m \rightarrow -B_m$ and
$J \rightarrow -J$.
The latter indicates that the entanglement is the same for the antiferromagnetic and ferromagnetic cases.  We thus restrict our considerations to
$B_m > 0$ and $J>0$.
Notice that the critical temperature
$T_{critical} \approx 1.13459 J$,
beyond which the thermal entanglement is zero, is independent of the magnetic field $B_m$.  This is in agreement with \cite{Wang}, where the concurrence \cite{Wootters, Hill} has been adopted as a measure of entanglement.\\

Now we look at Lee and Kim's two-qubit teleportation protocol, using two copies of the above two qubit thermal state, $\chi_{A_1B_1} \otimes \chi_{A_2B_2}$, as resource.  We consider as input two qubits in the Werner state \cite{Werner}
$
\rho_W = \frac{1}{4} (\sigma^0 \otimes \sigma^0 - \frac{2\Phi + 1}{3}
\sum_{i = 1}^3 \sigma^i \otimes \sigma^i)
$,
$
(-1 \leq \Phi \leq 1)
$.
The amount of entanglement associated with $\rho_W$ is given by
$
E(\rho_W) = \max\{\Phi, 0\}
$.
When $\Phi = 1$, $\rho_W = |\Psi^-\rangle\langle\Psi^-|$ is a maximally entangled pure state.  When $0 < \Phi < 1$, $\rho_W$ is an entangled mixed state.  Lastly, when $-1 \leq \Phi \leq 0$, $\rho_W$ is a separable mixed state.  Since our concern is entanglement teleportation, we focus on $0 < \Phi \leq 1$.
The output state is then given by \cite{Bose},
$$
\tilde{\rho}_W =
\Lambda(\chi_{A_1B_1} \otimes \chi_{A_2B_2}) \rho_W
$$
\begin{equation}
=
\sum_{j = 0}^3 
\sum_{k = 0}^3 tr\left[(E^j \otimes E^k)
(\chi_{A_1B_1} \otimes \chi_{A_2B_2})\right]
\left(\sigma^j_{A_1} \otimes \sigma^k_{A_2}\right) \rho_W
\left(\sigma^j_{A_1} \otimes \sigma^k_{A_2}\right)
\end{equation}
where
$E^0 = |\Psi^-\rangle \langle\Psi^-|$,
$E^1 = |\Phi^-\rangle \langle\Phi^-|$,
$E^2 = |\Phi^+\rangle \langle\Phi^+|$,
$E^3 = |\Psi^+\rangle \langle\Psi^+|$, and
$|\Phi^{\pm}\rangle = \frac{1}{\sqrt{2}} (|00\rangle + |11\rangle)$.\\

To characterize the quality of the teleported state $\tilde{\rho}_W$ it is often quite useful to look at the fidelity between $\rho_W$ and $\tilde{\rho}_W$, defined by \cite{Jozsa}
$$
F(\rho_W, \tilde{\rho}_W)
= \left\{
tr\left[
\sqrt{(\rho_W)^{\frac{1}{2}} \tilde{\rho}_W (\rho_W)^{\frac{1}{2}}}
\right]
\right\}^2
= \frac{1}{36(\cosh\beta B_m + \cosh\beta J)^2} \times
$$
$$
\left\{
2\sqrt{(1 - \Phi)\left[(1 - \Phi)\cosh^2\beta B_m +
2(2 + \Phi) \cosh\beta B_m \cosh\beta J +
(1 - \Phi)\cosh^2\beta J\right]}\right. +
$$
$$
\sqrt{(1 - \Phi)\left[(2 + \Phi) \cosh^2\beta B_m +
2 (1 - \Phi) \cosh\beta B_m \cosh\beta J +
(1 - \Phi) \cosh^2\beta J +
(1 + 2\Phi)\right]} +
$$
\begin{equation}
\left.
\sqrt{3}
\sqrt{(1 + \Phi)\left[(2 + \Phi) \cosh^2\beta B_m +
2 (1 - \Phi) \cosh\beta B_m \cosh\beta J +
3 (1 + \Phi) \cosh^2\beta J -
(1 + 2\Phi)\right]}
\right\}^2
\end{equation}
The concept of fidelity has been a useful indicator of the teleportation performance of a quantum channel when the input state is a pure state (see, e.g., \cite{Yeo} and references therein).  However it fails in our context where we consider the mixed Werner state.  In particular, we observe that, in the infinite temperature limit, $\beta \rightarrow 0$,  when there is zero thermal entanglement in the channels, we have
\begin{equation}
F(\rho_W, \tilde{\rho}_W) \rightarrow \frac{1}{4}\left[
(2 - \Phi) + \sqrt{3(1 - \Phi^2)}
\right]
\end{equation}
which increases as $\Phi \rightarrow 0$.\\

Returning to the main issue of this paper, we calculate the measure of entanglement for the teleported state $\tilde{\rho}_W$ to be
$$
E(\tilde{\rho}_W) =
\max\left\{
\frac{3\Phi\cosh^2\beta J - 2(2 + \Phi)\cosh\beta B_m \cosh\beta J - (1 - \Phi)\cosh^2\beta B_m - (1 + 2\Phi)}
{3(\cosh\beta B_m + \cosh\beta J)^2}
,0\right\}
$$
$$
= \max\left\{
\left[
\frac{(\cosh\beta B_m - \cosh\beta J)^2}{3(\cosh\beta B_m + \cosh\beta J)^2} +
\frac{2(\cosh^2\beta J - 1)}{3(\cosh\beta J + \cosh\beta B_m)^2}
\right] \Phi - \right.
$$
\begin{equation}
\left.
\frac{4\cosh\beta B_m \cosh\beta J + \cosh^2\beta B_m + 1}{3(\cosh\beta B_m + \cosh\beta J)^2}
,0\right\}
\end{equation}
In the zero temperature limit, $\beta \rightarrow \infty$, we have $E(\tilde{\rho}_W) \rightarrow \Phi$ when $B_m < J$, but $E(\tilde{\rho}_W) \rightarrow 0$ when $B_m \geq J$.  This is not difficult to understand when $B_m < J$, since at $T = 0$ the quantum channels are in the maximally entangled ground state:$\chi_{A_1B_1} = \chi_{A_2B_2} = |\Psi^-\rangle\langle\Psi^-|$ .  For $B_m > J$, the channels have zero entanglement and hence are not able to teleport entanglement.  However, at $B_m = J$, although the channels have nonzero entanglement, it does not allow them to perform better than ``classical channels'' \cite{Yeo}.  Since entanglement is a quantum property, we therefore do not expect the channels to teleport entanglement when $B_m = J$.\\

For nonzero temperatures, it is clear from (7) that entanglement is lost in the teleportation, since the coefficient of $\Phi$ is less than or equal to one, and the term independent of $\Phi$ is nonpositive.  The interesting thing to note is that the thermally entangled channels are still able to teleport some quantum entanglement even at nonzero temperatures.  In order to transmit nonzero entanglement, we require
\begin{equation}
\cosh\beta J > \frac{2 + \Phi}{3\Phi} \cosh\beta B_m +
\sqrt{\frac{-2\Phi^2 + 7\Phi + 4}{9\Phi^2}\cosh^2\beta B_m +
\frac{1 + 2\Phi}{3\Phi}}
\end{equation}
and hence $B_m < J$.  The `critical' temperature $T^{(m,\Phi)}_{critical}$ beyond which no quantum entanglement is teleported, is therefore dependent on the magnetic field $B_m$ and $\Phi$.  The right hand side of (8) is strictly greater than $\sqrt{2}$.  This shows that in order to teleport quantum entanglement, there must be some nonzero critical value of minimum thermal entanglement.   This minimum is not a constant but depends on $B_m$ and $T$ for some fixed $J$, in contrast to \cite{Lee}.  The right hand side of (8) increases as $\Phi\rightarrow 0$. This together with (7) indicate that quantum entanglement in a less entangled mixed state is more fragile to teleport.  It demands an even smaller $T_{critical}^{(m,\Phi)}$ for some fixed $B_m < J$, that is, $\exp(\beta J) \gg \exp(\beta B_m)$.\\

In conclusion, we have established the condition under which the quantum entanglement in a Werner state can be teleported via two separate, thermally entangled two-qubit Heisenberg XX chain.  We also demonstrated that the fidelity \cite{Jozsa} is in our case not a good indicator of the performance of these quantum channels.  Two qubit teleportation together with one-qubit unitary operations are sufficient to implement the universal gates for quantum computation \cite{Chuang}.  It is hoped that this paper would contribute to the gathering of phenomenology in this direction.\\

The author thank Yuri Suhov, Andrew Skeen and Suguru Furuta for useful discussions.  This publication is an output from project activity funded by The Cambridge MIT Institute Limited (``CMI'').  CMI is funded in part by the United Kingdom Government.  The activity was carried out for CMI by the University of Cambridge and Massachusetts Institute of Technology.  CMI can accept no responsibility for any information provided or views expressed.

\end{document}